\documentclass[useAMS,usenatbib,referee]{biom}

\newcommand{\logit}{\text{logit}}
\newcommand{\select}{\text{selected}}
\newcommand{\sigmai}{\sigma_i}
\newcommand{\sumk}{\sum_{k=1}^K}
\newcommand{\sumi}{\sum_{i=1}^{N}}
\newcommand{\E}{\mathbb E}
\newcommand{\R}{\mathbb R}
\newcommand{\p}{p}
\newcommand{\Rp}{\mathbb R_{+}}
\newcommand{\intP}{\int_0^{\infty}}
\newcommand{\intR}{\int_{-\infty}^{\infty}}
\newcommand{\byi}{\boldsymbol y_i}
\newcommand{\by}{\boldsymbol y}
\newcommand{\bz}{\boldsymbol z}
\newcommand{\bb}{\boldsymbol b}
\newcommand{\bSi}{\boldsymbol S_i}
\newcommand{\bSigmai}{\boldsymbol\Sigma_i}

\newcommand{\bsigma}{\boldsymbol\sigma}
\newcommand{\bSigma}{\boldsymbol\Sigma}
\newcommand{\bOmega}{\boldsymbol\Omega}
\newcommand{\with}{\text{ with }}

\newcommand{\btheta}{\boldsymbol \theta}

\newcommand{\SROC}{\text{SROC}}
\newcommand{\SAUC}{\text{SAUC}}

\newcommand{\sen}{\text{se}}
\newcommand{\spe}{\text{sp}}

\newcommand{\pmat}[4]{\begin{pmatrix} {{#1}} & {{#2}} \\ {{#3}} & {{#4}} \end{pmatrix}}
\newcommand{\svec}[2]{\left({{#1}},{{#2}}\right)^\top}
\newcommand{\bc}{\boldsymbol c}

\newcommand{\hSAUC}{\text{SA}\hat{\text{U}}\text{C}}

\usepackage{amsmath,amsfonts}
\usepackage{booktabs,multirow,tabularx}
\usepackage{url}
\usepackage{threeparttable}
\usepackage{color}
\usepackage{enumitem}   

\usepackage[figuresright]{rotating}

\title[Worst-case bounds for PB on the SROC curve]{Nonparametric worst-case bounds for publication bias on the summary receiver operating characteristic curve}

\author
{Yi Zhou\emailx{yzhou@pku.edu.cn} \\
Beijing International Center for Mathematical Research, Peking University, China
\and
Ao Huang\emailx{ao.huang@med.uni-goettingen.de} \\
Department of Medical Statistics, University Medical Center G{\"o}ttingen, Germany
\and
Satoshi Hattori\emailx{hattoris@biostat.med.osaka-u.ac.jp} \\
Department of Biomedical Statistics, Graduate School of Medicine;\\
Institute for Open and Transdisciplinary Research Initiatives, Osaka University, Japan}

\begin{document}



\pagerange{\pageref{firstpage}--\pageref{lastpage}} 
\volume{00}
\pubyear{2023}
\artmonth{12}

\doi{xx}

\label{firstpage}

\begin{abstract}
The summary receiver operating characteristic (SROC) curve has been recommended as one important meta-analytical summary to represent the accuracy of a diagnostic test in the presence of heterogeneous cutoff values.
However, selective publication of diagnostic studies for meta-analysis can induce publication bias (PB) on the estimate of the SROC curve. 
Several sensitivity analysis methods have been developed to quantify PB on the SROC curve, and all these methods utilize parametric selection functions to model the selective publication mechanism. 
The main contribution of this article is to propose a new sensitivity analysis approach that derives the worst-case bounds for the SROC curve by adopting nonparametric selection functions under minimal assumptions. 
The estimation procedures of the worst-case bounds use the Monte Carlo method to obtain the SROC curves along with the corresponding area under the curves in the worst case where the maximum possible PB under a range of marginal selection probabilities is considered.
We apply the proposed method to a real-world meta-analysis to show that the worst-case bounds of the SROC curves can provide useful insights for discussing the robustness of meta-analytical findings on diagnostic test accuracy. 
\end{abstract}


\maketitle

\section{Introduction}\label{sec:intro}

In studies that evaluate the capacity of biomarkers or diagnostic tests, the scientific question of interest is whether the biomarker or diagnostic test can accurately discriminate the diseased or non-diseased subjects. 
In clinical practice, diagnostic studies are conducted with limited numbers of subjects, so meta-analysis of diagnostic studies plays an important role in obtaining sound evidence on diagnostic test accuracy.
Sensitivity and specificity are widely accepted to quantify the diagnostic accuracy of binary diagnostic outcomes (positive or negative); thus, in studies that evaluate the diagnostic accuracy of a continuous biomarker, sensitivity and specificity are often reported by dichotomizing the measurements of the biomarker. 
Since sensitivity and specificity are usually estimated based on study-specific cutoff values, summarizing these cutoff-dependent measurements while ignoring the variation of cutoff values across multiple studies can lead to large heterogeneity in the meta-analytical results and difficulty in interpretations of those results.
In addition, sensitivities and specificities can be negatively correlated among studies.
Thus, it is not recommended to summarize sensitivities and specificities separately using the standard meta-analytical approach \citep{Macaskill2022}.  
As meta-analytical results of diagnostic test accuracy, the summary receiver operating characteristic (SROC) curve, as well as the area under the SROC curve (SAUC), are recommended by the Cochrane Diagnostic Test Accuracy Working Group \citep{Macaskill2022}.
One additional meta-analytical representation is the summary operating point (SOP), which is an integrated pair of sensitivity and specificity accounting for correlations of sensitivities and specificities among multiple studies. 
However, the representation of the SOP still relies on a certain cutoff value, which causes some difficulty in showing the diagnostic capacity under all possible cutoff values.
The SROC curve along with the SAUC and the SOP can be estimated by jointly modeling sensitivities and specificities using the bivariate normal model \citep{Reitsma2005} or the bivariate binomial model \citep{Rutter2001,Macaskill2004}.
Although the two models use different inference procedures and definitions of the SROC curves, they are regarded as different parametrizations of the same model without covariates \citep{Harbord2007}.

When conducting meta-analyses, one should pay attention to the presence of publication bias (PB) in the meta-analytical results. 
For univariate meta-analysis of intervention studies (e.g., randomized clinical trials) comparing treatment effects, people have found that larger studies and studies with significant results are more likely to be published; summarizing only the selectively published studies may cause biased estimates.
Many methods are available to identify or quantify PB in univariate meta-analysis. 
The funnel plot with regression-based \citep{Egger1997} or rank-based \citep{Begg1994} tests has been widely used to detect PB and the trim-and-fill method \citep{Duval2000a} to adjust for PB.  
All these methods detect PB by the asymmetry of the funnel plot.
However, selective publication may not be the only cause of the asymmetry of the funnel plot, and the results of these methods might be subjective and misleading. 
More delicate sensitivity analysis methods have developed, including the Copas-Heckman-type selection function \citep{Copas1999,Copas2000,Ning2017,Huang2021,Huang2023}, also known as the Heckman model \citep{Heckman1976,Heckman1979} in the field of econometrics, the $t$-statistic based selection function \citep{Copas2013}, and the worst-case bounds \citep{Copas2004}. 
A general review of these methods for univariate meta-analysis is given by \cite{Jin2015}.

In meta-analysis of diagnostic test accuracy, PB can also be a threat; however, methodological developments dealing with PB are limited. 
Detecting asymmetry of the funnel plot on the univariate and cutoff-independent diagnostic measurement such as the area under the receiver operating characteristic curves (AUC) would be one idea to address the PB issue. 
Since many diagnostic studies only report pairs of sensitivity and specificity without providing the AUC, it is impossible to apply the funnel plot, regression/rank-based method, or the trim-and-fill method.
Variations of the funnel plot on some univariate measurements such as sensitivity, specificity, and diagnostic odds ratio (DOR) have been proposed.
\cite{Song2002}, \cite{Deeks2005}, and \cite{Burkner2014} discussed the performance of funnel plot-based methods; however, their arguments are limited to detecting the existence of selective publication and cannot quantify the magnitude of potential bias on the SROC curve. 

In recent years, several methods have been proposed to quantify the impact of PB on the SROC curve.
\cite{Hattori2016} proposed the sensitivity analysis method based on the bivariate binomial model. 
\cite{Piao2019h} and \cite{Li2021} proposed the conditional and the empirical likelihood methods, respectively, based on the bivariate normal model to adjust PB. 
All these methods model the selective publication process by extending the Copas-Heckman-type selection function \citep{Copas1999,Copas2000}. 
Although the Copas-Heckman-type selection function describes selective publication processes to some extent, it could not explicitly show how the selection probability depends on the cutoff value of each study. 
\cite{Zhou2022} proposed a likelihood-based sensitivity analysis method by extending the idea of \cite{Copas2013}.
In their method, selection functions are proposed based on the $t$-type statistic of cutoff-dependent quantities such as sensitivity, specificity, or the DOR. 
The $t$-type statistic-based selection functions would be more appealing in modeling the selective publication mechanisms since the publication of each study can be influenced by its $P$-value or test statistic dependent of the cutoff value.

Despite of usefulness of these parametric selection functions in handling PB, it is still difficult to establish the explicit mechanism of selective publication in diagnostic studies due to the bivariate nature of sensitivity and specificity outcomes. 
Thus, methods relying on less explicit assumptions on the selective publication process would be more appealing. 
In this paper, we develop a nonparametric sensitivity analysis approach that relies on minimal assumptions to derive the worst-case bounds on the SROC curve. 
The proposed method extends the worst-case bounds approach of \cite{Copas2004} and constructs the bounds for the largest possible PB in estimating the SROC curve along with the SAUC.
For meta-analysis of intervention, \cite{Copas2004} proposed the theoretical bounds approach for PB over all the selection functions which was monotone with respect to the standard error (SE) of the outcome of each study; their approach derives the tight bounds under minimal assumptions, given a marginal selection probability.
Extending their method to meta-analysis of diagnostic test accuracy poses two challenges.
First, the estimand of interest, the SROC curve, is a non-linear formula of parameters to be estimated from the bivariate model. 
As the SROC curve can be viewed as the monotonic transformation of the linear combination of the logit-transformed integrated sensitivity and specificity, we aim to construct the worst-case bounds on the general linear combination of the logit-transformed sensitivity and specificity. 
The second challenge is the difficulty of extending the theory of \cite{Copas2004} into the bivariate model and deriving the theoretical bounds. 
We address this challenge by introducing a numerical approximation approach that solves the non-linear programming problem with linear constraints derived from the class of presumed selection functions. 
The proposed method is also applicable to meta-analysis of intervention and yields nearly identical results to theoretical bounds by \cite{Copas2004}.
In addition to bypassing the theoretical difficulties in deriving the worst-case bounds on the SROC curve, the flexibility of the numerical approximation approach enables us to consider several mechanisms of selective publication in a unified manner. 
We show that the proposed method can address the robustness of estimations regarding the SROC curve and the SAUC, providing valuable insights into the validity of meta-analyses of diagnostic test accuracy.

The remainder of this paper is organized as follows. 
In Section \ref{sec:unm}, we firstly review the random-effects model for the univariate meta-analysis and the worst-case bounds approach by \cite{Copas2004}. 
Subsequently, we introduce an alternative numerical approximation approach for estimating the worst-case bounds numerically.
In Section \ref{sec:diagmeta}, we introduce the bivariate normal model and the SROC curve in meta-analysis of diagnostic test accuracy and then propose the worst-case bounds and the estimation procedure. 
Each section includes an illustration of the proposed approach with a real-world meta-analysis example.
Finally, we conclude with a discussion in Section \ref{sec:dis}.

\section{Worst-case bounds in univariate meta-analysis of intervention}\label{sec:unm}

\subsection{Random-effects meta-analysis setting}\label{sec:model1}

Suppose that $N$ studies are published to evaluate the treatment effect (e.g., the log odds ratio). 
Each study $i$ provides a within-study variance $s_i^2$ and a point estimate $y_i$ of the true outcome valued $\theta_i$. 
Following the convention in the meta-analysis field, we regard $s_i$ as known and fixed. 
In the within-study level, the point estimate $y_i$ is regarded to have the following normal distribution:
\begin{align*}
y_i\mid\theta_i\sim N(\theta_i, s_i^2).
\end{align*}
The true values of $\theta_i$'s may vary across studies due to between-study heterogeneity (e.g., differences in populations, study designs, etc.).
It is assumed that $\theta_i$'s follow the normal distribution in the between-study level:
\begin{align*}
\theta_i\sim N(\theta, \tau^2),
\end{align*}
where $\theta$ is the overall mean of interest, and $\tau^2$ describes the between-study variance.
Marginally, $y_i$ has the following distribution:
\begin{align}\label{rem3}
y_i\sim N(\theta, \sigma_i^2)\with \sigma_i^2 = s_i^2 + \tau^2.
\end{align}
Allowing for the existence of between-study heterogeneity,
the inference of the unknown parameters ($\theta, \tau$) can be made based on the loglikelihood of model \eqref{rem3} by using the restricted or the ordinary maximum likelihood (ML) methods or other methods.
The ML estimation of $\theta$ is consistent with the restricted ML estimator, and in practice, the differences between the methods mainly exist in the estimation of $\tau$.
A general review of the estimation methods is given by \cite{Veroniki2016}.
In this paper, we adopt the ordinary ML method for estimation.

\subsection{Nonparametric selection functions and assumption}
\cite{Copas2004} proposed a nonparametric sensitivity analysis method that presents the worst-case bounds of PB in estimating the overall effect $\theta$ in model \eqref{rem3}.
Recall that each selected study in meta-analysis reports its own data $(y, \sigma)$.
In the absence of selective publication of studies, data $(y,\sigma)$ are regarded as random sample from the population of $S$ studies; then, $y$ and $\sigma$ have the following distributions:
\begin{equation}
\begin{aligned}\label{DofOP}
y \sim f(y\mid\sigma) = \dfrac{1}{\sigma}\phi\left(\dfrac{y-\theta}{\sigma}\right) 
\text{~and~}
\sigma\sim f(\sigma), 
\end{aligned}
\end{equation}
where $\phi(.)$ indicates the probability density function (pdf) of standard normal distribution, and $f(\sigma)$ is an unspecified pdf of $\sigma$, in which $\tau$ is regarded as estimated and fixed.
In meta-analysis, PB is usually modeled by a selective publication procedure where the published studies are regarded as selected ones and the unpublished studies the rejected ones. 
In the presence of selective publication, the selection of a study can be biased and is modeled by some selection function, denoted by
\begin{align}
p(y, \sigma) = P(\select  \mid  y, \sigma). \label{SFforU}
\end{align}
This selection function implies that the probability that each study is selected for meta-analysis depends on its outcome and the corresponding SE.
Selection function \eqref{SFforU} is not specified with any parametric model; however, the following assumption is made.
\begin{assumption}\label{CJAS}
$p(\sigma) = P(\select  \mid \sigma)$ is a non-increasing function of $\sigma$.
\end{assumption}
In Assumption \ref{CJAS}, selection function $p(\sigma)$ can be derived by $ p(\sigma) = \E\left\{p(y, \sigma)\mid\sigma\right\}$.
Assumption \ref{CJAS} implies that studies with smaller $\sigma$ (larger studies) are more likely to be selected than studies with larger $\sigma$ (smaller studies). 
This assumption is based on the empirical observation that small studies are more likely to be unpublished, which corresponds to the missingness of studies in the lower part of the funnel plot.

\subsection{Worst-case bounds for bias}

In the presence of selective publication, data $(y, \sigma)$ can be biased samples from pdfs \eqref{DofOP}.  
With selection functions $p(y, \sigma)$ and $p(\sigma)$, the joint distribution of $y$ and $\sigma$ conditional on a selected study is derived by
\begin{align*}
f(y, \sigma \mid \select) = \dfrac{1}{\p\sigma}\phi\left(\dfrac{y-\theta}{\sigma}\right)p(y,\sigma)f(\sigma), 
\end{align*}
where $\p$ is defined as the marginal (or overall) selection probability with $\p=P(\select)=\E\{p(\sigma)\}$. 
Let $b$ be the bias in the ML estimator of $\theta$, and then the bias can be solved by considering the expectation of the score function with respect to $\theta$:
\begin{align}
\E_O\{s(\theta + b)\} =\E\{s(\theta + b)\mid \select\} = 0,\label{EoS}
\end{align}
where $s(\theta)$ is the score function of model \eqref{rem3}, and $\E_O$ denotes the expectation of the selected (published) studies.
By introducing $z = \sigma^{-1}(y-\theta)$, which follows the standard normal distribution, $N(0,1)$, the bias $b$ is derived by
\begin{align}\label{b1}
b = \left\{\intP\sigma^{-2}p(\sigma)f(\sigma)d\sigma\right\}^{-1}{\intP\intR\sigma^{-1}z\phi(z)p(\sigma z+\theta, \sigma)f(\sigma)dzd\sigma}.
\end{align}

Given the monotonicity of $p(\sigma)$ in Assumption (\ref{CJAS}) and a specific value for $\p$,
\cite{Copas2004} derived the upper and lower bounds to the bias $b$ in their Theorem 2, that is,
\begin{align*}
|b| \le \frac{\E_O(\sigma^{-1})}{\E_O(\sigma^{-2})}\dfrac{\phi\left\{\Phi^{-1}(\p)\right\}}{\p},
\end{align*}
where $\Phi(.)$ denotes the standard normal cumulative distribution function and $\E_O$ the expectation of the observed $\sigma$.
The lower and upper bounds for $b$ are attained by certain selection functions of step functional forms.
Expectations $\E_O(\sigma^{-1})$ and $\E_O(\sigma^{-2})$ can be estimated by the empirical means of $\sigmai^{-1}=(s_i^2+\tau^2)^{-1/2}$ and $\sigmai^{-2}=(s_i^2+\tau^2)^{-1}$ over the published studies, respectively. 
\cite{Copas2004} proposed to conduct the sensitivity analysis by calculating the empirical bounds of $b$ (hereinafter, the Copas-Jackson bound), denoted by $b_\text{CJ}$, that is,
\begin{align}\label{emp_bound}
b_\text{CJ}= \dfrac{\sumi\sigmai^{-1}}{\sumi\sigmai^{-2}}\frac{\phi\left\{\Phi^{-1}(\p)\right\}}{\p}
\text{ with }-b_\text{CJ}\le b \le b_\text{CJ}.
\end{align}
The Copas-Jackson bound is a function of $\p$, which can be approximately represented by $\p\approx N/S$. 
In practice, $\p$ is unknown and then regarded as a sensitivity parameter.  
By taking a plausible range of values of $\p$, the Copas-Jackson bound gives the maximum bias (in estimating $\theta$) in terms of $\p$ as well as the number of unpublished studies (i.e., $S-N$).

Recall that $\sigmai^2=s_i^2+\tau^2$ in model \eqref{rem3}, and we regard $\tau^2$ as fixed when deriving the bounds. 
One way to fix $\tau^2$ is to replace it with its estimator, such as its ML estimator, by assuming that the selective publication process does not affect the estimation of between-study variance. 
On the other hand, as done in the example of \cite{Copas2004}, some plausible values were set by looking at the confidence interval (CI) of $\tau^2$.      

\subsection{Simulation-based worst-case bounds for bias}\label{sec:uni-sim}

Alternative to the empirical Copas-Jackson bound \eqref{emp_bound}, we propose an approximation approach to obtain the lower and upper bounds of $b$ numerically.
Consider the pdf of the observed $\sigma$ conditional on a selected study, derived by
\begin{align*}
f(\sigma\mid\select) = \dfrac{P(\select\mid\sigma)f(\sigma)}{P(\select)} = \dfrac{p(\sigma)f(\sigma)}{\p}.
\end{align*}
Under this formulation, the asymptotic bias $b$ \eqref{b1} can be written by
\begin{align}
b &= \left\{\intP\sigma^{-2}\p f(\sigma\mid\select)d\sigma\right\}^{-1}{\intP\intR\sigma^{-1}z\phi(z)p(\sigma z+\theta, \sigma)\dfrac{\p f(\sigma\mid\select)}{p(\sigma)}dzd\sigma}\nonumber\\
&=\left[\E_O\left\{\sigma^{-2}\right\}\right]^{-1}\E_O\left[\dfrac{\sigma^{-1}}{p(\sigma)}\E_z\left\{z p(\sigma z+\theta, \sigma)\right\}\right]\label{bEo}.
\end{align}
Recall that $z \sim N(0,1)$. 
By simulating a sequence of $K$ random variables $(z_1, z_2,\dots, z_K)$ from $N(0,1)$, 
$\E_z\left\{z p(\sigma z+\theta, \sigma)\right\}$ is approximated by $K^{-1}\sumk {z_k}p(\sigma_i^{-1}z_k+\theta, \sigmai)$ using the Monte Carlo method.
Then, the bias $b$ \eqref{bEo} is approximated by $b_\text{MC}$, that is,
\begin{align*}
b_\text{MC}=\left\{
\sumi\sigmai^{-2}
\right\}^{-1}
\left\{\dfrac{1}{K}\sumi\sumk \sigmai^{-1}p_i^{-1}{z_k}p_{i,k}
\right\},
\end{align*}
where we denote $p_{i,k}= p(\sigma_i^{-1}z_k+\theta, \sigmai)$ and $p_{i} = p(\sigmai)$, and they are unknown parameters.
By calculating the maximum and minimum values of $b_\text{MC}$ under some constraints on $p_{i,k}$ and $p_{i}$, we construct the simulation-based worst-case bounds for bias. 
This is achieved through the solution of the nonlinear programming problem:
\begin{center}
maximize or minimize $b_\text{MC}$ 
\end{center}
\begin{equation*}
\begin{aligned}
    \text{subject to: } & \text{(C1) } 0\le p_{i,k}, p_i \le 1~(1\le i\le N, 0 \le k\le K); \\ 
    &\text{(C2) } p_i= \dfrac{1}{K}\sum_{k=1}^K p_{i,k}; \text{(C3) }  \dfrac{1}{N}\sumi p_{i}\ge \p; 
    \text{(C4) } p_i\le p_{i+1} \text{ when } \sigmai \ge \sigma_{i+1}.
\end{aligned}
\end{equation*}

Condition (C1) indicates the the probabilistic properties of $p(y,\sigma)$ and $p(\sigma)$.
Condition (C2) holds when $K$ is large since $p(\sigma)=\E\left[p(y,\sigma)\mid\sigma\right]$.
Condition (C3) holds because $\p = \E \{p(\sigma)\} \approx S^{-1}\sum_{i=1}^S p_{i} \le N^{-1}\sum_{i=1}^N p_{i}$, where the number of population studies ($S$) is no smaller than the number of published studies $(N)$. 
Condition (C4) is resulted from the monotonicity of $p(\sigma)$ in Assumption \ref{CJAS}.
Notably, the marginal selection probability $\p$ in Condition (C3) should be prespecified by plausible values for sensitivity analysis.

The minimum and maximum values of $b_\text{MC}$ under Conditions (C1)-(C4) can be solved by nonlinear optimization methods such as the interior point method, which is conducted by \texttt{OPTMODEL} Procedure in \texttt{SAS} (version 9.4) in this paper.

\subsection{Example 1}\label{eg1}

To validate the performance of simulation-based bounds, we re-visit the illustrative meta-analysis in \cite{Copas2004}.
This is a meta-analysis of 14 randomized clinical trials concerning whether the use of prophylactic corticosteroids could improve the survival of premature infants. 
The events are the death of infants, and the detailed data of the treatment and control groups are summarized in Table 1 of \cite{Copas2004}.
The parameter of interest is the overall log odds ratio (lnOR).
Since the ML estimate of $\tau^2$ is 0, we set $\sigma_i^2 = s_i^2$.
Without considering selective publication, the overall lnOR was estimated to be $-0.480$ with 95\% CI to be $[-0.707, -0.244]$, indicating that the treatment was significantly effective.

\begin{figure}
\begin{center}
\centerline{\includegraphics[width=\textwidth]{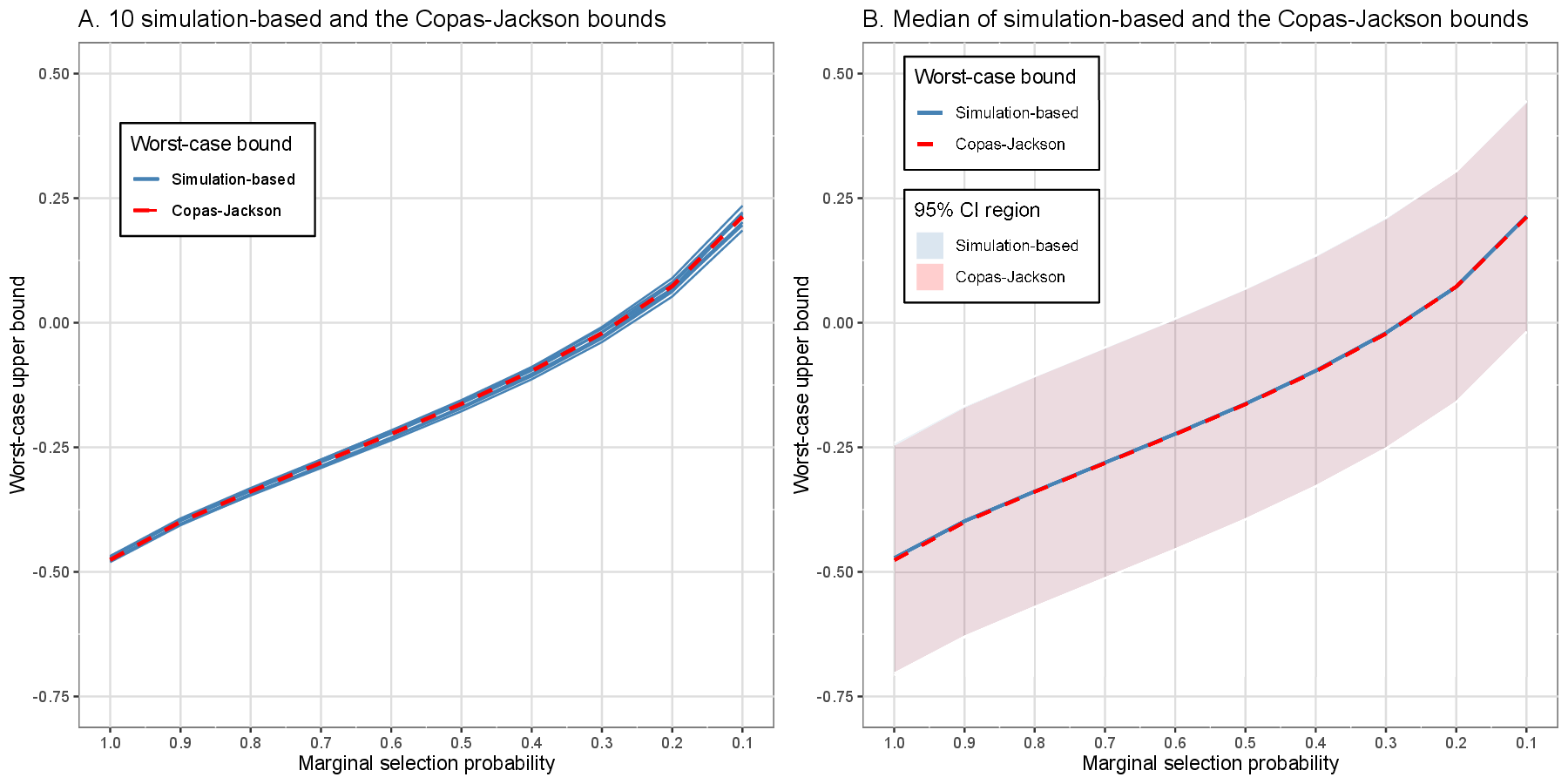}}
\end{center}
\caption{Example 1: the upper bounds of the lnOR given by the simulation-based bounds and the Copas-Jackson bound. \label{fig:eg1}}
\end{figure}

Since we were interested in the upper bounds of the overall lnOR taking into account a range of marginal selection probabilities, we estimated the maximum values of $b_\text{MC}$ and $b_\text{CJ}$ given $\p=1,0.9,\dots, 0.1$, which is equivalent to the number of potentially unpublished studies increasing from 0 to 126.
For the simulation-based bounds, we generated $K=2000$ independent random variables $z_k$ from $N(0,1)$.
To assess the influence of the random number generation for $z$, we computed the simulation-based upper bounds 10 times.
The comparison between 10 simulation-based upper bounds and the upper bound of $b_\text{CJ}$ on the lnOR is presented in Figure \ref{fig:eg1}A; 
the comparison between the median of simulation-based bounds and the upper bound of $b_\text{CJ}$ is presented in Figure \ref{fig:eg1}B.
In Figure \ref{fig:eg1}B, the 95\% CIs of the bounds were added by crudely adopting the SE without accounting for PB. 
The corresponding values are summarized in Table \ref{tab:tab1}.

\begin{table}

\caption{\label{tab:tab1}Example 1: the upper bounds of the lnOR with 95\% CI given by the median of 10 simulation-based bounds and the Copas-Jackson bound.}
\centering
\begin{tabular}[t]{rrr}
\toprule
$\p$ & Simulation-based bound [95\% CI] & Copas-Jackson bound [95\% CI]\\
\midrule
1.0 & -0.473 [-0.705, -0.242] & -0.476 [-0.707, -0.244]\\
0.9 & -0.399 [-0.631, -0.168] & -0.399 [-0.631, -0.168]\\
0.8 & -0.339 [-0.571, -0.108] & -0.339 [-0.570, -0.107]\\
0.7 & -0.282 [-0.514, -0.051] & -0.281 [-0.512, -0.050]\\
0.6 & -0.225 [-0.456, 0.007] & -0.223 [-0.455, 0.008]\\
0.5 & -0.164 [-0.396, 0.067] & -0.163 [-0.394, 0.068]\\
0.4 & -0.098 [-0.330, 0.133] & -0.097 [-0.328, 0.134]\\
0.3 & -0.022 [-0.254, 0.209] & -0.021 [-0.253, 0.210]\\
0.2 & 0.072 [-0.159, 0.303] & 0.073 [-0.158, 0.305]\\
0.1 & 0.211 [-0.021, 0.442] & 0.213 [-0.019, 0.444]\\
\bottomrule
\end{tabular}
\end{table}

The simulation-based bounds showed stable performance in Figure \ref{fig:eg1}A and gave nearly identical results as the Copas-Jackson bound in Figure \ref{fig:eg1}B. 
The medians of the simulation-based bounds gave very close estimates to the Copas-Jackson bound. 
When $p = 0.6$, assuming that about $(14/0.6)-14=9$ studies (40\% of studies) were unpublished from the population, the 95\% CI of the bias-adjusted lnOR began to cover 0.
When more than 60\% of studies (9 studies) were assumed to be unpublished, one could doubt the significance of the overall lnOR in meta-analysis.
This example demonstrates that the numerical approximation approach yields desirable results in estimating the worst-case bounds for the lnOR, which inspired us to extend this approach to address the PB issue in more complex meta-analysis of diagnostic test accuracy.

\section{Worst-case bounds in meta-analysis of diagnostic test accuracy and example}\label{sec:diagmeta}

\subsection{Bivariate random-effects model and the SROC curve}

In meta-analysis of diagnostic test accuracy, each study $i~(i = 1, 2, \dots, N$) provides the observed numbers of true positives, false negatives, true negatives, and false positives, denoted by $n_{11,i}, n_{01,i}, n_{00,i}$, and $n_{10,i}$, respectively. 
Let $n_{+1,i} = n_{11,i} + n_{01,i}$ and $n_{+0,i} =n_{00,i} + n_{10,i}$ be the numbers of diseased and non-diseased subjects, respectively.
These data can be succinctly summarized in a confusion matrix, as illustrated in Table \ref{tab:confusion}.

\begin{center}
\begin{table}[!hbt]
\caption{Notations of cell frequencies in $2 \times 2$ confusion matrix for diagnostic study $i$.
\label{tab:confusion}}
\centering
\begin{tabular*}{0.5\linewidth}{@{\extracolsep\fill}rrrr@{\extracolsep\fill}}
\toprule
\multicolumn{2}{l}{\multirow{2}{*}} & \multicolumn{2}{c}{Truth} \\ \cline{3-4} 
\multicolumn{2}{l}{}                  & Diseased  	   & Non-diseased  \\ 
\midrule
Test        & Positive     & $n_{11,i}$ & $n_{10,i}$\\
\multirow{2}{*}{}      & Negative     & $n_{01,i}$ & $n_{00,i}$\\
\addlinespace
                       & Total        & $n_{+1,i}$ & $n_{+0,i}$\\ 
\bottomrule                       
\end{tabular*}
\end{table}
\end{center}

When frequencies of zero are found in Table \ref{tab:confusion}, the continuity correction is often applied by adding 0.5 to all the cells.
Sensitivity and specificity in each study are estimated by $\sen_i = {n_{11,i}}/{n_{+1,i}}$ and $\spe_i = {n_{00,i}}/{n_{+0,i}}$, respectively.
In practice, sensitivity and specificity are reported in pairs, and these pairs should be synthesized by the bivariate model.
\cite{Reitsma2005} proposed the bivariate normal model (hereinafter, the Reitsma model) to model the logit-transformed sensitivity and specificity pairs.
This model can be regarded as the bivariate extension of the random-effects model introduced in Section \ref{sec:model1}.
Let $y_{1,i}$ and $y_{2,i}$ be the logit-transformed $\sen_i$ and $\spe_i$, respectively, where the logit-transformation is defined by $\logit(x) = \log(x)- \log(1-x)$.
Correspondingly, $s_{1,i}^2$ and $s_{2,i}^2$ represent the variances of $y_{1,i}$ and $y_{2,i}$.
When $n_{+1,i}$ and $n_{+0,i}$ are large and $\sen_i$ and $\spe_i$ are neither close to 0 nor 1, the variances can be estimated by $s_{1,i}^2 = 1/n_{11,i}+1/n_{01,i}$ and $s_{2,i}^2=1/n_{01,i}+1/n_{00,i}$, respectively.

The Reitsma model assumes that $\svec{y_{1,i}}{y_{2,i}}$ has the bivariate normal distribution in the within-study level:
\begin{align*}
\binom{y_{1,i}}{y_{2,i}}\left|\binom{\theta_{1,i}}{\theta_{2,i}}\right.\sim N \left(\binom{\theta_{1,i}}{\theta_{2,i}}, \bSi\right)
\with \bSi = \pmat{s_{1,i}^2}{0}{0}{s_{2,i}^2},
\end{align*}
where $\svec{\theta_{1,i}}{\theta_{2,i}}$ is the vector of the true logit-transformed sensitivity and specificity pair for study $i$.
In the between-study level, $\svec{\theta_{1,i}}{\theta_{2,i}}$ is assumed to follow the normal distribution:
\begin{align*}
\binom{\theta_{1,i}}{\theta_{2,i}}\sim N \left(\binom{\theta_1}{\theta_2}, \bOmega\right)
\with \bOmega = \pmat{\tau_{1}^2}{\tau_{12}}{\tau_{12}}{\tau_{2}^2},
\end{align*}
where $(\theta_1,\theta_2)^\top$ is the vector of overall true logit-transformed sensitivity and specificity pair and $\bOmega$ the between-study covariance matrix; 
$\tau_1^2$ is the variance of $\theta_{1,i}$, $\tau_2^2$ is the variance of $\theta_{2,i}$, and $\tau_{12}=\rho\tau_1\tau_2$ is the covariance between $\theta_{1,i}$ and $\theta_{2,i}$;
$-1\le \rho\le 1$ is the correlation coefficient.
Denote $\byi = \svec{y_{1,i}}{y_{2,i}}$ and $\btheta = (\theta_1,\theta_2)^\top$, the Reitsma model gives the marginal distribution of $\byi$, that is,
\begin{align}\label{reitsma}
\byi \sim N(\btheta, \bSigmai), \with \bSigmai = \bSi+\bOmega.
\end{align}
The unknown parameters $\btheta$ and $\boldsymbol \Omega$ can be estimated by the ML method.
(Although different ML estimation methods can be employed, the estimates of $\btheta$ are consistent in different methods and the differences exist in the estimates of $\bOmega$.)
Applying the inverse of logit-transformation to $\btheta$ yields the SOP, i.e., 
$\svec{\sen}{\spe} = \svec{\logit^{-1}(\theta_1)}{\logit^{-1}(\theta_2)}$.

Accounting for the heterogeneous cutoff values in diagnostic studies, the SROC curve and the SAUC are important meta-analytical summaries of test accuracy.
The SROC curve is derived by the conditional expectation of $\theta_{1,i}$ given $\theta_{2,i}$.
Specifically, letting $x$ be $1-\spe$, the SROC curve and the SAUC are defined as follows \citep{Reitsma2005}:
\begin{align*}
\SROC(x; \btheta, \bOmega) 
&= \logit^{-1} \left[ \theta_1 - \dfrac{\tau_{12}}{\tau_2^2}\{\logit(x)+\theta_2\} \right]
\\
\SAUC(\btheta, \bOmega) 
&= \int_{0}^{1}\SROC(x; \btheta, \bOmega)dx.
\end{align*}
The SROC curve and the SAUC can be estimated by replacing the unknown theoretical quantities with their ML estimators.

\subsection{Nonparametric selection functions and assumptions}

We express the covariance matrix in the Reitsma model \eqref{reitsma} as
\begin{align*}
\bSigma = \pmat{\sigma_{1}^2}{\tau_{12}}{\tau_{12}}{\sigma_{2}^2}
\text{ with } \sigma_{1}^2= {s_{1}^2+\tau_1^2} \text{ and } \sigma_{2}^2= {s_{2}^2+\tau_2^2}
\end{align*}
Similar to the practice in univariate meta-analysis, we tentatively estimate and fix $(\tau_1, \tau_{12}, \tau_2)$ to derive the worst-case bounds.  
Let $\bsigma = \svec{\sigma_{1}}{\sigma_{2}}$ be the vector of observed SEs with fixed $\tau_1$ and $\tau_2$. 
Without selective publication, the observed data $(\by, \bsigma)$ from multiple diagnostic studies are regarded as random samples from the population with the following pdfs:
\begin{align*}
\by \sim f(\by\mid \bsigma)=|\bSigma|^{-1/2}\phi_2\left\{\bSigma^{-1/2}(\by-\btheta)\right\} \text{~and~}
\bsigma \sim f(\bsigma) = f(\sigma_1, \sigma_2),
\end{align*}
where
$\phi_2(.)$ denotes the pdf of bivariate standard normal distribution,
and $f(\bsigma)$ is an unspecified pdf for $\sigma_{1}$ and $\sigma_{2}$.
Under selective publication, the selection function \eqref{SFforU} is extended as
\begin{align*}
p(\by, \bsigma) = P(\select \mid \by, \bsigma), 
\end{align*}
which models the probability that a diagnostic study is selected for meta-analysis conditional on its logit-transformed sensitivity and specificity and the corresponding SEs.
%

Due to the bivariate nature of outcomes, the publication process in diagnosis studies should be more complicated compared to univariate meta-analysis.
By applying the funnel plot-based methods to certain univariate measurements, such as the logit-transformed sensitivity, specificity, and the lnDOR, asymmetry of the funnel plot on that measurement may be detected, raising concerns about the selective publication process influenced by that particular measurement\citep{Deeks2005, Burkner2014}.
This motivates us to explore the impact of PB by assuming different selective publication processes.
%
Thus, we consider the general situation and model the selective publication using the linear combination of $\sigma_1^2$ and $\sigma_2^2$, which is formally stated in the following assumption.
\begin{assumption}\label{AS2}
$p(\bsigma) = P(\select\mid \sigma_1, \sigma_2)=\E\left[p(\by, \bsigma)\mid\bsigma\right]$ is a non-increasing function of $\beta_1\sigma_{1}^2+\beta_2\sigma_{2}^2$, and, without loss of generality, $0 \le \beta_1;\beta_2\le 1$.
\end{assumption}
This assumption allows one to examine the impact of bias conditional on various selective publication mechanisms by specifying values of $(\beta_1,\beta_2)$ as part of sensitivity analysis.
To implement the monotonicity of $p(\bsigma)$ in practice, one could consider the following particular situations:
\begin{enumerate}[label=(\roman*), labelwidth=2em,itemindent=!]
\item Let $p(\bsigma)$ be a non-increasing function of $\sigma_1^2$ (i.e., $\beta_1=1, \beta_2=0$); this allows for the evaluation of bias impact under the assumption that a study with smaller $\sigma_{1}$ (indicating more significant sensitivity) is more likely to be selected.
\item Let $p(\bsigma)$ be a non-increasing function of $\sigma_2^2$ (i.e., $\beta_1=0, \beta_2=1$), which assumes a study with more significant specificity is more likely to be selected.
\item Let $p(\bsigma)$ be a non-increasing function of $\sigma_1^2+\sigma_2^2$ (i.e., $\beta_1=1, \beta_2=1$), which assumes a study with more significant lnDOR is more likely to be selected.
\end{enumerate}

\subsection{Worst-case bounds for bias vector}
Under selective publication, the joint distribution of $\by$ and $\bsigma$ for a selected diagnostic study is derived by
\begin{align*}
f(\by, \bsigma\mid\select) &= \dfrac{P(\select\mid\by, \bsigma)f(\by \mid \bsigma)f(\bsigma)}{P(\select)}\\
&=\dfrac{1}{\check p|\bSigma|^{1/2}}\phi_2\left\{\bSigma^{-1/2}(\by-\btheta)\right\}p(\by,\bsigma)f(\bsigma),
\end{align*}
where $\p=\E\{p(\bsigma)\}$ is the marginal selection probability for diagnostic studies.
Under selective publication, data $(\by, \bsigma)$ are regarded as biased sample from the Reitsma model. 
Let $\bb = (b_1, b_2)^\top$ be the bias vector in estimating $\btheta$. 
The arguments for the univariate case in \eqref{EoS} are extended as follows. The bias vector $\bb$ satisfies the equation
\begin{align*}
\E_O\{s(\btheta + \bb)\} = \E\{s(\btheta + \bb)\mid \select\} =\int_{\Rp^2}\int_{\R^2} s(\btheta+\bb) f(\by, \bsigma\mid\select) d\by d\bsigma = 0,
\end{align*}
where $s(\btheta)=\bSigma^{-1}(\by-\btheta)$ is the score function of $\btheta$ from the Reitsma model;
$\bsigma\in \Rp^2=\{(\sigma_1,\sigma_2)\mid 0<\sigma_1,\sigma_2<\infty\}$, 
and $\by\in \R^2=\{(y_1,y_2)\mid-\infty<y_1,y_2<\infty\}$.
Let $\bz = \bSigma^{-1/2}\left(\by-\btheta\right)$ following the bivariate standard normal distribution; then, the bias vector is derived by
\begin{align}\label{b21}
\bb &=\left\{
\int_{\Rp^2}
\bSigma^{-1}p(\bsigma) f(\bsigma)d\bsigma\right\}^{-1}
\left\{\int_{\Rp^2}\int_{\R^2}\bSigma^{-1/2}\bz\phi_2(\bz)f(\bsigma)
p(\bSigma^{1/2}\bz +\btheta, \bsigma) d\bz d\bsigma\right\}
\end{align}
The detailed derivation is presented in Web Appendix A.

Let $f_O(\bsigma)$ be the pdf of the observed $\bsigma$ conditional on the selected (published) studies; we can derive $f_O(\bsigma)$ as
\begin{align*}
f_O(\bsigma)&= f(\bsigma\mid\select)
= \dfrac{P(\select\mid\bsigma)f(\bsigma)}{P(\select)}
=\dfrac{p(\bsigma)f(\bsigma)}{\p}.
\end{align*}
Thus, the bias vector $\bb$ \eqref{b21} is re-expressed by the following equation:
\begin{align}
\bb &=\left\{
\int_{\mathbb R_+^2}
\boldsymbol\Sigma^{-1}
\p f_O(\boldsymbol \sigma)
d\boldsymbol\sigma
\right\}^{-1}
\left\{
\int_{\mathbb{R}_+^2} \boldsymbol\Sigma^{-1/2} \left\{\int_{\mathbb{R}^2}\boldsymbol z \phi_2(\bz)p(\boldsymbol \Sigma^{1/2} \boldsymbol z+\boldsymbol\mu,\boldsymbol \sigma)d \boldsymbol z \right\}{\dfrac{\p f_O(\boldsymbol \sigma)}{p(\boldsymbol \sigma)}}d\boldsymbol\sigma
\right\}\nonumber\\
&=\left \{\E_O\left [\bSigma^{-1}\right ] \right \}^{-1}
\E_O\left \{ \dfrac{\bSigma^{-1/2}}{{p(\bsigma)}}\E_{\bz} \left [\bz {p(\bSigma^{-1/2} \bz+\btheta,\bsigma)}  \right]\right \}.\label{b2}
\end{align}

\subsection{Simulation-based worst-case bounds for bias vector}

Following the idea in Section \ref{sec:uni-sim} for the univariate meta-analysis, we propose to use the Monte Carlo method to obtain the worst-case bounds for the bias vector $\bb$.
Let $\boldsymbol z_k = \svec{z_{1,k}}{z_{2,k}}~(k = 1,2,\dots, K)$ be the random vectors sampled from the bivariate standard normal distribution.
When $K$ is large, Equation \eqref{b2} is approximated by
\begin{align}\label{MCB2}
\bb_\text{MC} =\binom{b_{1,\text{MC}}}{b_{2,\text{MC}}}=\left\{\sumi\bSigma_i^{-1}\right\}^{-1}
\dfrac{1}{K}\sumi\sumk\left\{\bSigmai^{-1/2}\boldsymbol z_k
\dfrac{\tilde p_{i,k}}{\tilde p_i}\right\},
\end{align}
where $\tilde p_{i,k}=p(\bSigmai^{-1/2} \bz_k+\btheta,\bsigma_i)$ and $\tilde p_{i}=p(\bsigma_i)$ are the unknown parameters.
To obtain the extreme values of each element in $\bb_\text{MC}$, we introduce the contrast vector, denoted by $\bc = \svec{c_1}{c_2}\in\R^2$. 
In general, we consider the worst-case bounds of the linear combination $\bc^\top \bb_\text{MC}=c_1b_{1,\text{MC}}+c_2b_{2,\text{MC}}$.
Specifically, when $\bc=(1,0)$ or $(0,1)$, $\bb_\text{MC}$ provides the worst-case bounds of $b_{1,\text{MC}}$ on $\theta_1$ or those of $b_{2,\text{MC}}$ on $\theta_2$, respectively.
Given marginal selection probability $\p$, the simulation-based worst-case bounds of $\bc^\top\bb_\text{MC}$ can be solved by the following nonlinear programming problem under some constraints of $\tilde p_{i,k}$ and $\tilde p_{i}$:
\begin{center}
maximize or minimize $\bc^\top \bb_\text{MC}$ 
\end{center}
\begin{equation}
\begin{aligned}
    \text{subject to: } & \text{(D1) } 0\le \tilde p_{i,k}, \tilde p_i \le 1~(1\le i\le N, 0 \le k\le K); \\ 
    &\text{(D2) } \tilde p_i= K^{-1}\sum_{k=1}^K \tilde p_{i,k}; \text{(D3) }  N^{-1}\sumi \tilde p_{i}\ge \p; 
    \text{(D4) Assumption \ref{AS2}}.\label{nlp2}
\end{aligned}
\end{equation}

To realize (D4) Assumption \ref{AS2}, we may consider one of the following constraints, which correspond to (i) to (iii), respectively:
\begin{enumerate}[labelwidth=2em,itemindent=!]
\item [(D4.1)] $\tilde p_{i}\le \tilde p_{i+1}$ when $\sigma_{1,i}^2\ge \sigma_{1,i+1}^2$, assuming that $p(\bsigma)$ is a non-increasing function of $\sigma_1^2$;
\item [(D4.2)] $\tilde p_{i}\le \tilde p_{i+1}$ when $\sigma_{2,i}^2\ge \sigma_{2,i+1}^2$, assuming that $p(\bsigma)$ is a non-increasing function of $\sigma_2^2$;
\item [(D4.3)] $\tilde p_{i}\le \tilde p_{i+1}$ when $\sigma_{1,i}^2+\sigma_{2,i}^2\ge \sigma_{1,i+1}^2+\sigma_{2,i+1}^2$, assuming that $p(\bsigma)$ is a non-increasing function of $\sigma_1^2+\sigma_2^2$.
\end{enumerate}

On the other hand, the SROC curve is the monotonic inverse logit transformation of the linear combination of $\theta_1$ and $\theta_2$, which enables the consideration of the largest bias on the SROC curve as well as the SAUC. 
The SROC curve taking into account the bias vector $\bb$ is derived by
\begin{align}\label{srocb}
\SROC(x;\btheta+\bb, \bOmega)&=\logit^{-1}\left[
(\theta_1+b_1)-\dfrac{\tau_{12}}{\tau_2^2}\left\{\logit(x)+(\theta_2+b_2)\right\}\right]\nonumber\\
&= \logit^{-1}\left[\theta_1-\dfrac{\tau_{12}}{\tau_2^2}\left\{\logit(x)+\theta_2\right\} +\left(1,-\dfrac{\tau_{12}}{\tau_2^2}\right)\binom{b_1}{b_2}\right]\nonumber\\
&= \logit^{-1}\left[\theta_1-\dfrac{\tau_{12}}{\tau_2^2}\left\{\logit(x)+\theta_2\right\} +\tilde \bc^\top\bb\right].
\end{align}
By fixing $\tau_1$, $\tau_2$ and $\tau_{12}$ as sensitivity parameters and taking $\tilde{\bc}=(1, -\tau_{12}/\tau_2^2)^\top$ as the contrast vector, the worst-case bounds of the SROC curve established based on the minimum and the maximum values of $\tilde\bc^\top\bb_\text{MC}$.
These extreme values can be solved via the nonlinear programming problem under constraints \eqref{nlp2}.
Let $\tilde b_\text{MC}^L$ and $\tilde b_\text{MC}^U$ be the minimum and the maximum values of $\tilde\bc^\top\bb_\text{MC}$, bounding $\tilde\bc^\top\bb_\text{MC}$ within $[\tilde b_\text{MC}^L,\tilde b_\text{MC}^U]$.
The SAUC accounting for bias is derived by integrating \eqref{srocb}, and its worst-case bounds are derived by $[\SAUC_\text{MC}^L, \SAUC_\text{MC}^U]$ with
\begin{align*}
&\SAUC_\text{MC}^L = \int_0^1\logit^{-1}\left[\theta_1-\dfrac{\tau_{12}}{\tau_2^2}\left\{\logit(x)+\theta_2\right\} +\tilde b_\text{MC}^L\right]dx\\
&\SAUC_\text{MC}^U = \int_0^1\logit^{-1}\left[\theta_1-\dfrac{\tau_{12}}{\tau_2^2}\left\{\logit(x)+\theta_2\right\} +\tilde b_\text{MC}^U\right]dx.
\end{align*}

In previous formulas, we treated $(\tau_1, \tau_{12}, \tau_2)$ as sensitivity parameters and fixed; however, they are unknown in practice.
One solution is to replace $(\tau_1, \tau_{12}, \tau_2)$ with their ML estimators by assuming that selective publication would not affect the estimation of between-study variances or some other plausible values as demonstrated by \cite{Copas2004}.

\subsection{Example 2}

We demonstrate the application of simulation-based worst-case bounds for the SROC curve and the SAUC by re-analyzing the meta-analysis for the prognostic value of troponins in acute pulmonary embolism \citep{Becattini2007}.
This meta-analysis investigated the relationship between troponin and short-term death and included 20 studies with binary outcomes.
Since it had the same data structure as meta-analysis of diagnostic test accuracy, we analyzed this meta-analysis using the SROC curve \citep{Reitsma2005}.
The data are presented in Web Appendix B.
For studies with zero entries, we conducted continuity correction by adding 0.5 to all the cells of those studies.
Without considering selective publication, the Reitsma model estimated the SAUC with its 95\% CI to be $0.724~[0.639, 0.795]$.
The ML estimates of the between-study variances were $(\hat\tau_1^2,\hat\tau_{12},\hat\tau_2^2) =(0.171, -0.283, 0.588)$.

To explore the impact of PB in the meta-analytical results, we constructed the lower bounds on the SROC curve given decreasing marginal selection probabilities ($\p$) and different selective publication processes according to Conditions (D4.1)-(D4.3).
In the computation of simulation-based bounds $\bb_\text{MC}$, we generated 2000 random vectors $\bz$.
As done for the univariate meta-analysis in Section \ref{eg1}, we applied the proposed methods with 10 different sets of $\bz$ to assess the potential influence of random number generation. 
Figure \ref{fig:eg2-1}(A)-(C) illustrates the 10-time estimates of lower bounds on the SROC curve given $\p=1, 0.8, \dots, 0.2$ under Conditions (D4.1)-(D4.3), respectively; Figure \ref{fig:eg2-1}(D) presents the 10-time estimates of the bounds on the SAUC given $ \p=1, 0.9, \dots, 0.1$.
\begin{figure}
\begin{center}
\centerline{\includegraphics[width=\textwidth]{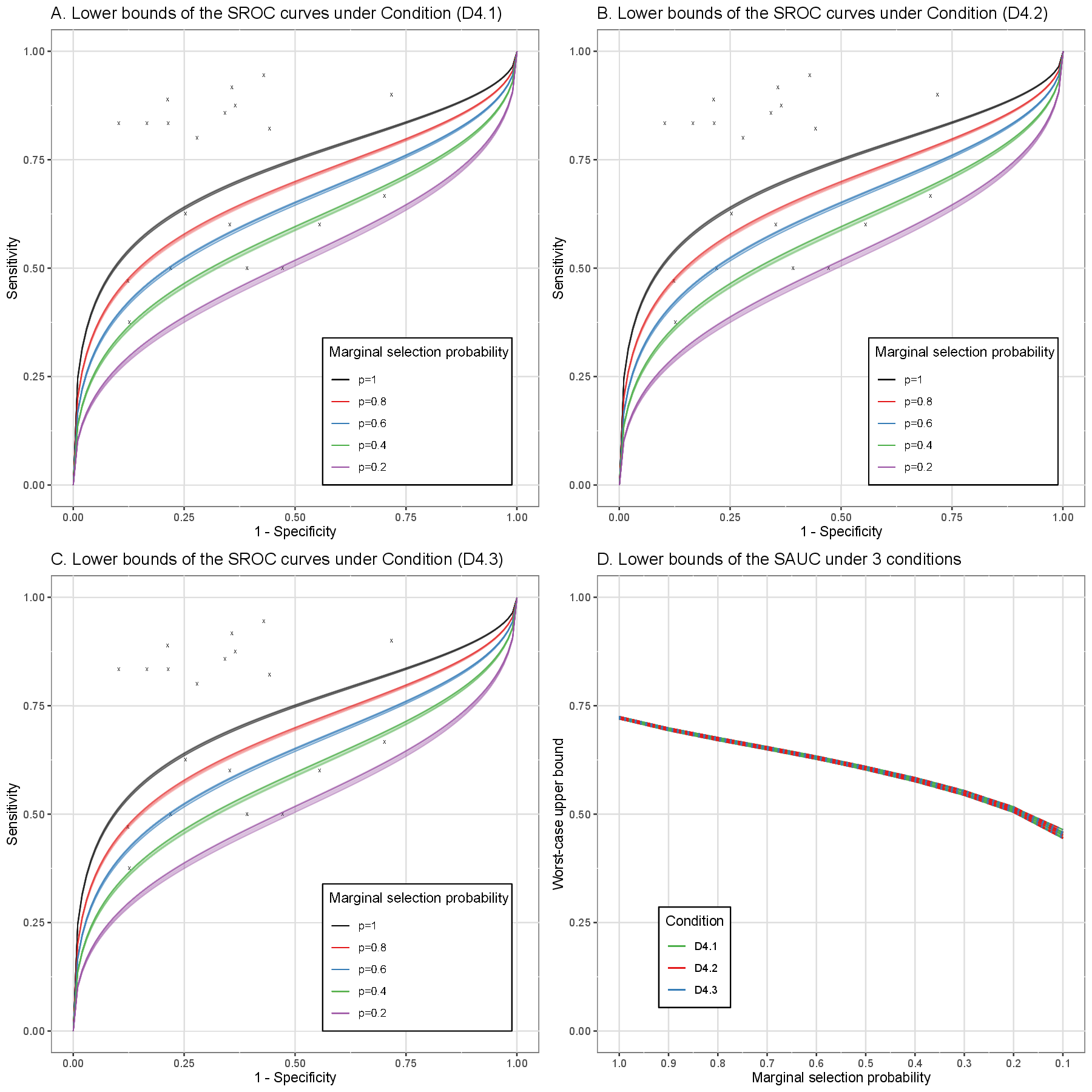}}
\end{center}
\caption{Example 2: 10-time estimates of the simulation-based worst-case bounds on the SROC curves and the corresponding SAUC under Conditions (D4.1)-(D4.3), respectively. \label{fig:eg2-1}}
\end{figure}

The 10-time simulation-based bounds showed nearly same estimates under Conditions (D4.1)-(D4.3) with minimal variations.
Taking the medians among the 10 estimates, we established the worst-case lower bounds on the SROC curve and the SAUC.
The corresponding results are shown in Figure \ref{fig:eg2-2};
and in Figure \ref{fig:eg2-2}(D), we presented 95\% CI for the lower bounds on the SAUC, however, using the SEs of parameters in the absence of PB. 
The detailed calculation of CIs is presented in Web Appendix C.

\begin{figure}
\begin{center}
\centerline{\includegraphics[width=\textwidth]{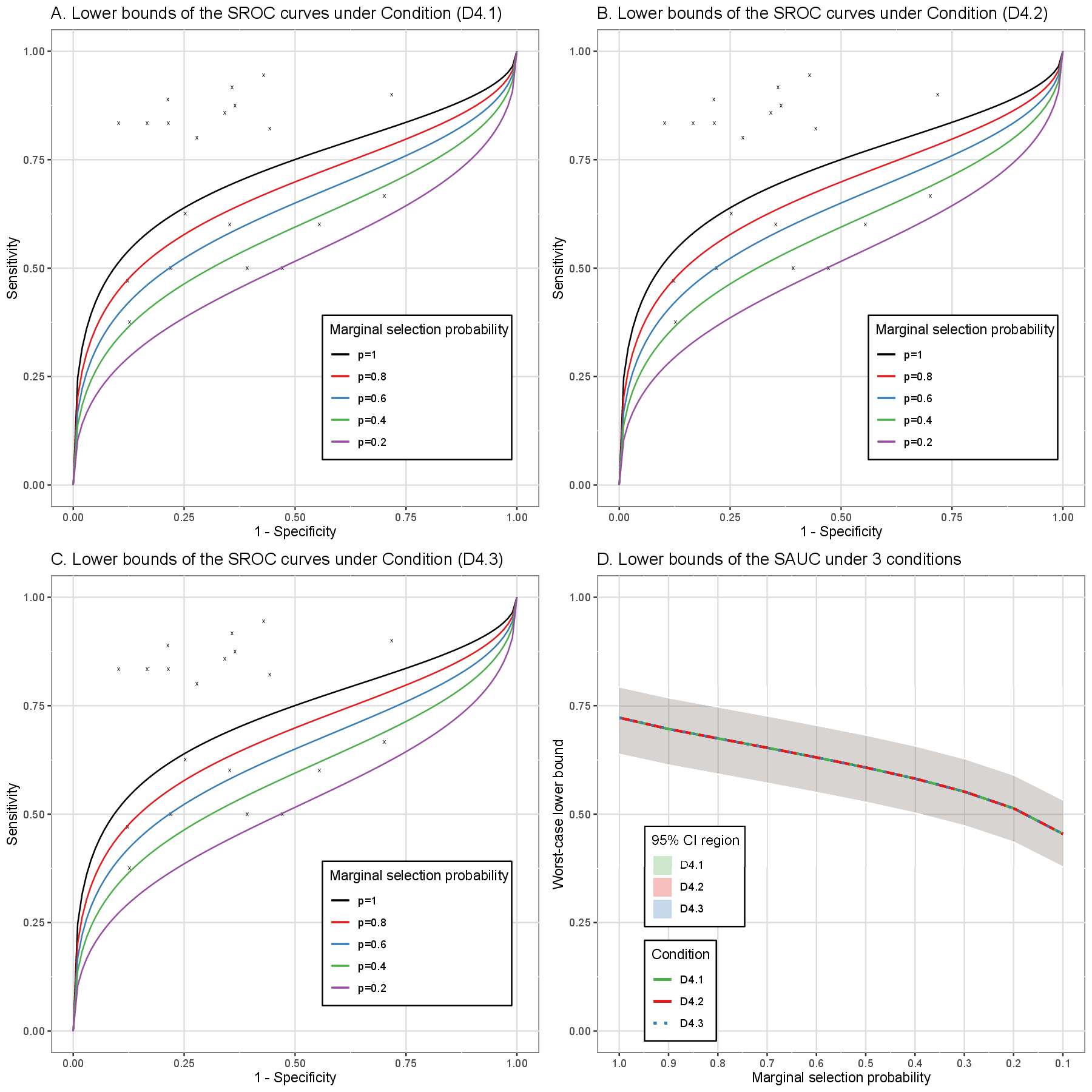}}
\end{center}
\caption{Example 2: median of 10-time estimates of the simulation-based worst-case bounds on the SROC curves and the corresponding SAUC under Conditions (D4.1)-(D4.3), respectively \label{fig:eg2-2}}
\end{figure}

When $\p=1$, the SROC curves and the SAUCs were estimated in the absence of selective publication.
When $\p>0.4$, that is, more than 30 studies were potentially unpublished, the CI of the SAUC tended to cover 0.5, indicating the nonsignificant test accuracy.
The estimates of worst-case bounds of the SAUC were summarized in Table \ref{tab:tab2}.

\begin{table}
\caption{\label{tab:tab2}Example 2: median of the simulation-based worst-case bounds on the SAUC.}
\centering
\begin{tabular}[t]{rr}
\toprule
$\p$ & Lower bounds on the SAUC [95\% CI] \\
\midrule
1.0 & 0.723 [0.638, 0.794]\\
0.9 & 0.697 [0.614, 0.769]\\
0.8 & 0.675 [0.593, 0.748]\\
0.7 & 0.653 [0.572, 0.727]\\
0.6 & 0.631 [0.550, 0.705]\\
0.5 & 0.608 [0.528, 0.683]\\
0.4 & 0.582 [0.503, 0.658]\\
0.3 & 0.552 [0.474, 0.628]\\
0.2 & 0.514 [0.436, 0.591]\\
0.1 & 0.455 [0.378, 0.533]\\
\bottomrule
\end{tabular}
\end{table}
 
When $\p$ was as small as 0.1, the SAUC decreased to 0.455 .
When no more than 50\% of studies were assumed to be unpublished (i.e., $\p\ge 0.5$), the worst-case bounds of the SAUC maintained above 0.5, indicating the good diagnostic test accuracy of troponin in acute pulmonary embolism.
In some meta-analyses, the worst-case bounds can be estimated inconsistently under certain conditions.
We presented another example for illustration in Web Appendix D.

\section{Conclusions}\label{sec:dis}

In this paper, we developed a simulation-based method to construct the worst-case bounds for quantifying the impact of potential PB on the SROC curve and the SAUC in meta-analysis of diagnostic test accuracy.
By calculating the bounds given fixed values of the marginal selection probability, one can argue whether the diagnostic capacity suggested by the SROC curves is sufficiently stable against potential unpublished studies. 
Since it is difficult to assume explicit selective publication processes in diagnostic studies, it would be valuable to develop bounds under minimal assumptions of selection functions; 
%
As a nonparametric sensitivity analysis method, the proposed method does not rely on any specific parametric assumptions.
The selection functions in the proposed method rely on the minimal assumption that the selection function given the variances of outcomes is the non-increasing function.
Due to the bivariate nature of outcomes in diagnostic studies, we recommend conducting sensitivity analyses against various scenarios of the selective publication process, even if similar results may be achieved.
The proposed method provided one more option for evaluating the robustness of the SROC curve and the SAUC results for meta-analysis of diagnostic test accuracy. 
%

One novelty of the proposed method is to make it possible to construct the worst-case bounds for a complicated quantity of the SROC curve.
The idea was to translate the problem into the construction of bounds for the linear combination of bias vector $\bc^\top \bb$ with $\bc=(1, -\tau_{12}/\tau_2^2)$, regarding $\tau_{12}/\tau_2^2$  as a sensitivity parameter.   
Technically, the worst-case bounds on the integrated sensitivity, specificity, or the lnDOR can be constructed by specifying $\bc=(1,0), (0,1)$, or $(1,1)$, respectively.
However, these quantities depend on various cutoff values of studies collected for meta-analysis, and it is not easy to interpret the diagnostic test accuracy using these quantities.
As mentioned, the success of our development was due to regarding the heterogeneity parameters $(\tau_1, \tau_{12}, \tau_2)$ as sensitivity parameters. 
It is also a limitation of our method; we need to set relevant values for these sensitivity parameters. 
A crude approach was to use the ML estimate ($\hat{\tau}_1^2$, $\hat{\tau}_{12}$, $\hat{\tau}_2^2$) by assuming that selective publication did not affect the estimation of the between-study covariance matrix. 
It is recommended to evaluate its influence by changing some plausible values.

In recent years, sensitivity analysis methods for PB in meta-analysis of intervention have been extended to multivariate meta-analysis. 
The Copas-Heckman-type selection function \citep{Copas1999,Copas2000} has been extended to meta-analysis of multiple treatments comparison \citep{Chootrakool2011}, network meta-analysis \citep{Mavridis2013,Mavridis2014,MarksAnglin2021}, and meta-analysis of diagnostic studies \citep{Hattori2018,Piao2019h,Li2021}. 
Some extensions of the $t$-statistic based selection function \citep{Copas2013} were proposed for meta-analyses of diagnostic studies \citep{Zhou2022} or prognosis studies \citep{Zhou2023}. 
To the best of our knowledge, the worst-case bounds approach has not been introduced to any multivariate meta-analyses and our method should be the first proposal.   
The proposed method with the estimation procedure can be easily extended for evaluating the impact of PB in other bivariate or multivariate meta-analyses based on the normal random-effects model, as long as the assumptions are plausible enough.
To interpret the results of worst-case bounds, some reference intervals such as CIs would be very helpful. 
In our examples, CIs of the bounds were calculated informally by using the SEs estimated without accounting for any selective publication. 
For meta-analysis of intervention, \cite{Henmi2007} developed a kind of worst-case CI accounting for the selective publication process. 
It would be useful to extend this methodology to meta-analysis of diagnostic test accuracy.  
%

\section{Software}
\label{sec5}

Software in the form of R and SAS codes, together with a sample input data set and complete documentation is available on GitHub at \url{https://github.com/meta2020/worstcase1-codes}.

\backmatter





\bibliographystyle{biom} \bibliography{ref_bib}

\section*{Supporting Information}
Web Appendices, referenced in Sections 3.3 and 3.5, are available with this paper at the Biometrics website on Wiley Online Library.
\vspace*{-8pt}

\label{lastpage}

\section*{Supporting information for ``Nonparametric worst-case bounds for publication bias on the summary receiver operating characteristic curve''}

\section*{Web Appendix A}

As mentioned in Section 3.3 after Equation (9), we present the derivation of the asymptotic bias vector $\bb$.
Denote $\bb = (b_1, b_2)^\top$ to be the asymptotic bias in estimating $\btheta$, and then $\bb$ is derived by the solution to
\begin{align*}
\E_O\{s(\btheta + \bb)\} =\int_{\Rp^2}\int_{\R^2} s(\btheta+\bb) f_O(\by, \bsigma) d\by d\bsigma = 0
\end{align*}

Because $s(\btheta+\bb)=\bSigma^{-1}\{\by-\btheta-\bb\}$ and we denote $f_O(\by, \bsigma)=f(\by, \bsigma\mid \select)=\dfrac{|\bSigma|^{-1/2}}{\p}\phi_2\left\{\bSigma^{-1/2}(\by-\btheta)\right\}p(\by,\bsigma)f(\bsigma)$, we can obtain

\begin{align*}
\E_O\{s(\btheta + \bb)\}=\int_{\Rp^2}\int_{\R^2} \bSigma^{-1}\{\by-\btheta-\bb\} \dfrac{|\bSigma|^{-1/2}}{\p}\phi_2\left\{\bSigma^{-1/2}(\by-\btheta)\right\}p(\by,\bsigma)f(\bsigma) d\by d\bsigma=0
\end{align*}

Equivalently,
\begin{align*}
&\int_{\Rp^2}\int_{\R^2} \bSigma^{-1}\{\by-\btheta\} \dfrac{|\bSigma|^{-1/2}}{\p}\phi_2\left\{\bSigma^{-1/2}(\by-\btheta)\right\}p(\by,\bsigma)f(\bsigma) d\by d\bsigma\\
&=\int_{\Rp^2}\int_{\R^2} \bSigma^{-1}\bb \dfrac{|\bSigma|^{-1/2}}{\p}\phi_2\left\{\bSigma^{-1/2}(\by-\btheta)\right\}p(\by,\bsigma)f(\bsigma) d\by d\bsigma.
\end{align*}

Let $\bz = \bSigma^{-1/2}(\by-\btheta)$, then we obtain
\begin{align}
&\int_{\Rp^2}\int_{\R^2} \bSigma^{-1/2}\bz \phi_2\{\bz\}p(\bSigma^{1/2}\bz+\btheta,\bsigma)f(\bsigma) d\bz d\bsigma\label{eq1}\\
&=\int_{\Rp^2}\int_{\R^2} \bSigma^{-1}\bb \phi_2\{\bz\}p(\bSigma^{1/2}\bz+\btheta,\bsigma)f(\bsigma) d\bz d\bsigma.\label{eq2}
\end{align}

Since 
\begin{align*}
p(\bsigma) &= P(\select\mid \sigma_1, \sigma_2)\\
&=\E\left[p(\by, \bsigma)\mid\bsigma\right]\\
&=\int_{\R^2}p(\by, \bsigma)f(\by\mid\bsigma)d\by\\
&=\int_{\R^2}p(\by, \bsigma)|\bSigma|^{-1/2}\phi_2\left\{\bSigma^{-1/2}(\by-\btheta)\right\}d\by\\
&= \int_{\R^2}p(\bSigma^{1/2}\bz+\btheta, \bsigma)\phi_2\{\bz\}d\bz,
\end{align*}

Equation \eqref{eq2} can be written as
\begin{align*}
&\int_{\Rp^2}\int_{\R^2} \bSigma^{-1}\bb \phi_2\{\bz\}p(\bSigma^{1/2}\bz+\btheta,\bsigma)f(\bsigma) d\bz d\bsigma\\
&=\int_{\Rp^2}\bSigma^{-1} p(\bsigma)f(\bsigma)  d\bsigma \cdot\bb
\end{align*}

Thus, equation \eqref{eq1} is expressed as
\begin{align*}
&\int_{\Rp^2}\int_{\R^2} \bSigma^{-1/2}\bz \phi_2\{\bz\}p(\bSigma^{1/2}\bz+\btheta,\bsigma)f(\bsigma) d\bz d\bsigma\\
&=\int_{\Rp^2}\bSigma^{-1} p(\bsigma)f(\bsigma)  d\bsigma \cdot\bb.
\end{align*}

Finally, the asymptotic bias $\bb$ is derived by 
\begin{align*}
\bb=&\left\{\int_{\Rp^2}\bSigma^{-1} p(\bsigma)f(\bsigma)  d\bsigma\right\}^{-1}\left\{\int_{\Rp^2}\int_{\R^2} \bSigma^{-1/2}\bz \phi_2(\bz)p(\bSigma^{1/2}\bz+\btheta,\bsigma)f(\bsigma) d\bz d\bsigma\right\}
\end{align*}

\section*{Web Appendix B}

As mentioned in Section 3.5, we present the data of Example 2.

\begin{table}[]
\caption{List of troponin data}
\label{tab:trop}
\centering
\begin{tabular}{@{}rrrrr@{}}
\toprule
study & TP & FP & FN & TN  \\ \midrule
1     & 0  & 14 & 0  & 22  \\
2     & 0  & 5  & 0  & 19  \\
3     & 1  & 17 & 1  & 19  \\
4     & 4  & 24 & 1  & 62  \\
5     & 5  & 9  & 1  & 33  \\
6     & 8  & 16 & 9  & 114 \\
7     & 23 & 50 & 5  & 63  \\
8     & 4  & 16 & 0  & 6   \\
9     & 6  & 40 & 1  & 77  \\
10    & 6  & 56 & 10 & 386 \\
11    & 2  & 40 & 1  & 17  \\
12    & 12 & 50 & 8  & 40  \\
13    & 8  & 24 & 0  & 32  \\
14    & 8  & 10 & 1  & 37  \\
15    & 5  & 36 & 0  & 65  \\
16    & 7  & 19 & 1  & 33  \\
17    & 9  & 30 & 6  & 55  \\
18    & 10 & 18 & 6  & 53  \\
19    & 2  & 4  & 0  & 22  \\
20    & 2  & 18 & 0  & 161 \\ \bottomrule
\end{tabular}%
\end{table}

\section*{Web Appendix C}

The confidence interval (CI) of the SAUC is calculated by using the delta method. 
We define 
$$\boldsymbol{D} = \int_{0}^{1}\SROC(x;\btheta, \bOmega)\{1-\SROC(x;\btheta, \bOmega)\}\nabla \SROC(x;\btheta, \bOmega) dx,$$
where $\nabla \SROC(x)$ is the gradient of the SROC curve and 
$$\nabla \SROC(x)=(1,-\rho\tau_1/\tau_2, -\rho/\tau_2\{\mathrm{logit}(x)+\mu_2\}, \rho\tau_1/\tau_2^2\{\mathrm{logit}(x)+\mu_2\}, -\tau_1/\tau_2\{\mathrm{logit}(x)+\mu_2\})^\top.$$
Then,
$\hat{\boldsymbol{D}}$ denotes the $\boldsymbol{D}$ with its unknown parameters replaced with their ML estimates without accounting for publication bias, that is,
$$\hat{\boldsymbol{D}} = \int_{0}^{1}\SROC(x;\hat\btheta+\bb_\text{MC}, \hat\bOmega)\{1-\SROC(x;\hat\btheta+\bb_\text{MC}, \hat\bOmega)\}\nabla \SROC(x;\hat\btheta+\bb_\text{MC}, \hat\bOmega) dx,$$
Denote $\hat{\boldsymbol{\Sigma}}$ to be the estimated variance-covariance matrix of $(\hat{\boldsymbol{\mu}}, \hat{\boldsymbol{\Omega}})$ from the ML method in the absence of publication bias.
According to the delta method, the variance of the SAUC is consistently estimated by 
$$\mathbb V[\hSAUC] = \hat{\boldsymbol{D}}^T \hat{\boldsymbol{\Sigma}} \hat{\boldsymbol{D}}.$$
By applying the delta method to logit-transformed $\hSAUC$, a two-tailed CI of $\hSAUC$ at significant level $\eta$ can be estimated by
\begin{align*}
\mathrm{logit}^{-1} \left \{
\mathrm{logit} (\hSAUC) \pm  z_{1-{\eta}/2}
\dfrac{\sqrt{\mathbb V[\hSAUC]}}{\hSAUC(1-\hSAUC)} \right \}.
\end{align*}


\section*{Web Appendix D}

\cite{Safdar2005} conducted the meta-analysis to identify the most accurate method for diagnosing intravascular device (IVD) related bloodstream infection. 
In their meta-analysis, 33 diagnostic studies of semi-quantitative and quantitative catheter segment culture tests were analyzed.
For studies with zero entries, the continuity correction was conducted by adding 0.5 to the data of those studies.
Without considering selective publication, the Reitsma model was conducted and the SAUC was estimated to be $0.873$ with the 95\% CI to be $[0.805, 0.919]$.
The between-study variance were estimated to be $\hat\tau_1^2 = 0.357$, $\hat\tau_{12} = -0.209$, and $\hat\tau_2^2 = 0.683$.

Following the same analytical procedure in Example 2, we 
We calculated the lower bounds of the SROC curve and the SAUC given decreasing values of $\p$ and under Conditions (D4.1)-(D4.3)
In calculating $\bb_\text{MC}$, we generated 2000 random vectors $\bz$ for the Monte Carlo approximation and repeated the calculation 10 times.
Figure \ref{fig:eg2-1}A-C presented 10 simulated worst-case bounds of the SROC constrained by Conditions (C1)-(C3) and additional Condition (D1), (D2), or (D3), respectively.
Figure \ref{fig:eg2-1}D presented the corresponding SAUC.
Figure \ref{fig:eg2-2}(A)-(D) presented the corresponding medians.

The values of the worst-case bounds are presented in Table \ref{tab:tab2}. 

\begin{figure}
\begin{center}
\centerline{\includegraphics[width=\textwidth]{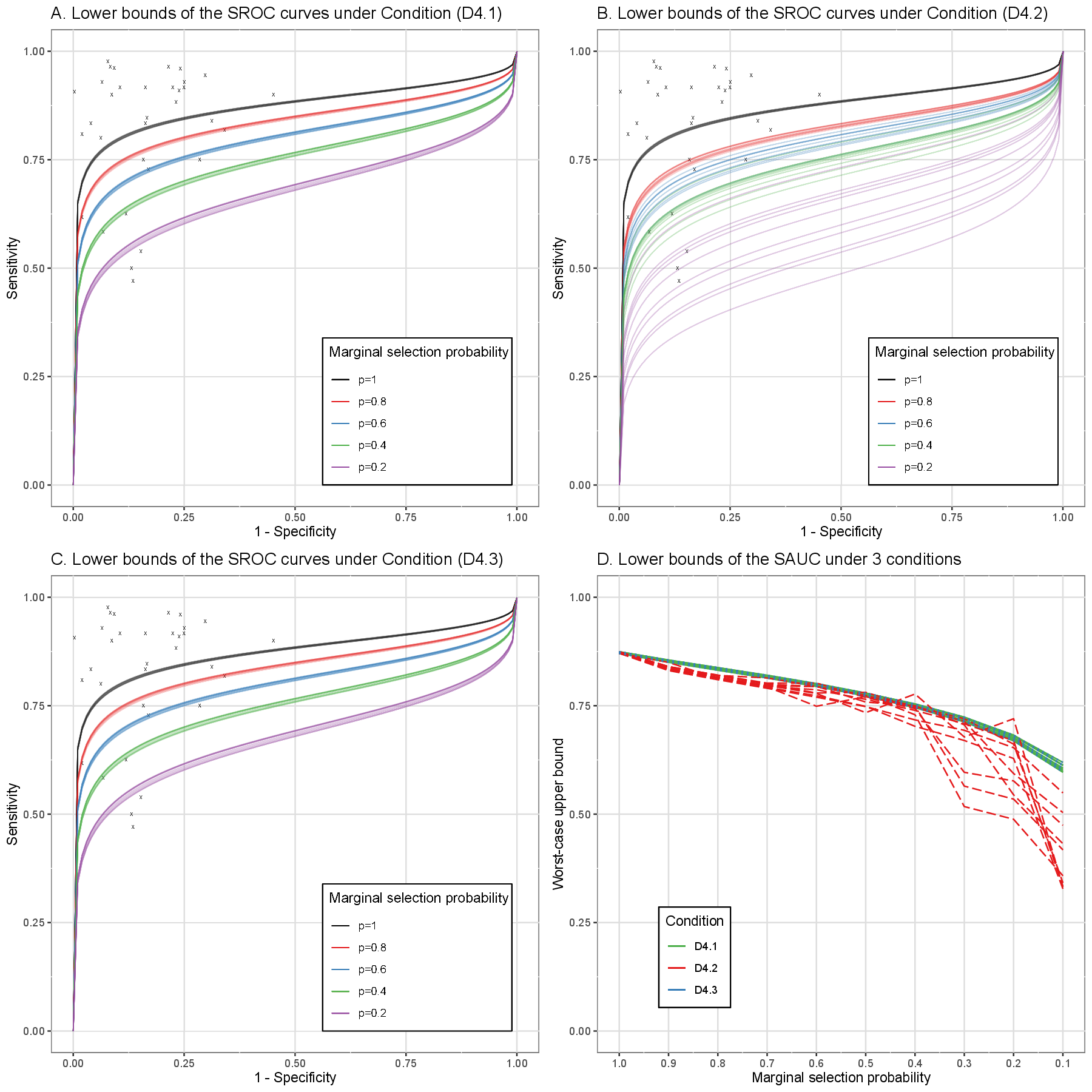}}
\end{center}
\caption{Example 2: the changes of the SROC curve under the assumption of Condition (D1)-(D3). \label{fig:eg2-1}}
\end{figure}

\begin{figure}
\begin{center}
\centerline{\includegraphics[width=\textwidth]{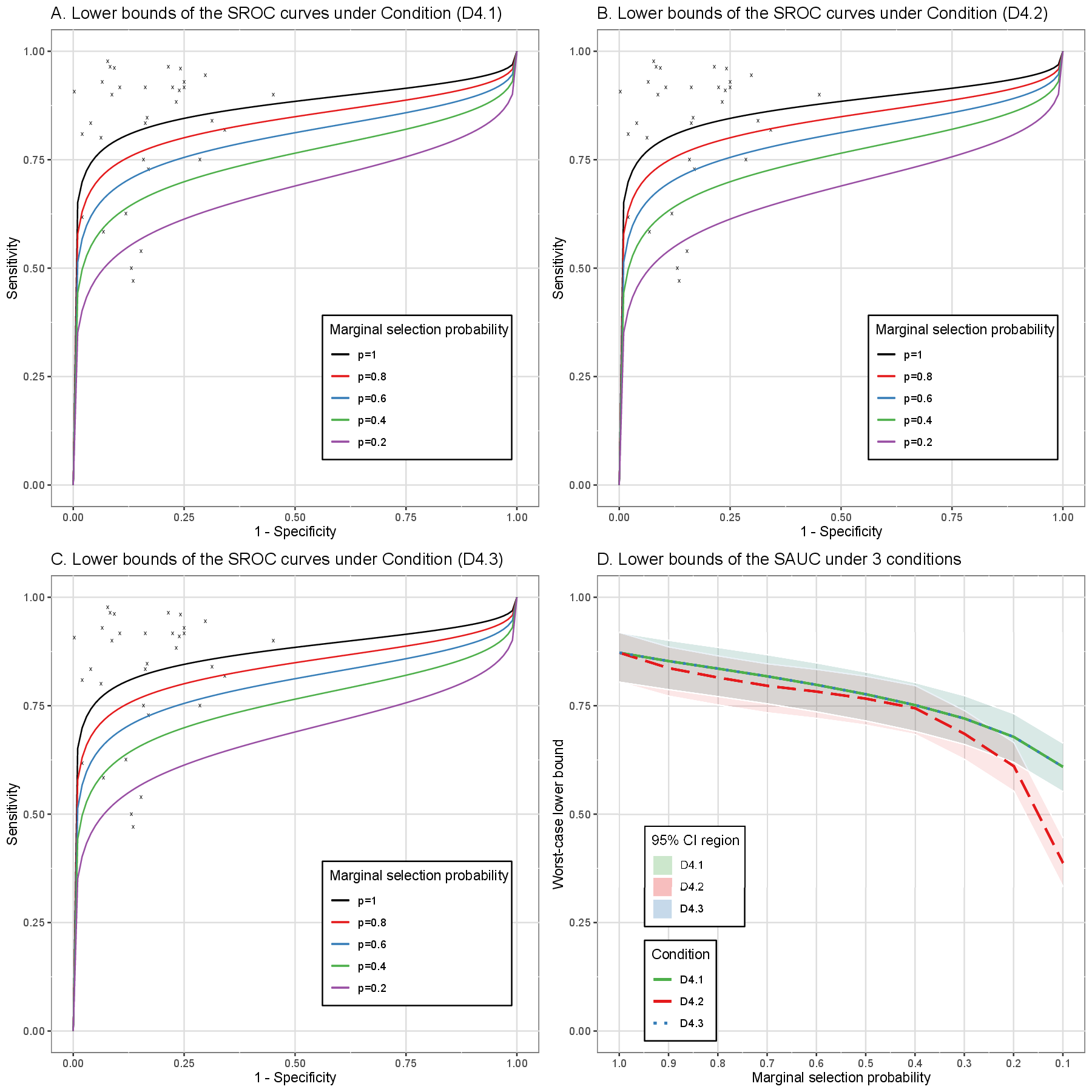}}
\end{center}
\caption{Example 2 (average over 10 repeats): the changes of the SROC curve under the assumption of Condition (D1)-(D3). \label{fig:eg2-2}}
\end{figure}

\begin{table}
\caption{\label{tab:tab2}Example 2: the lower worst-case bounds of the SAUC under the assumption of Condition (D1)-(D3) with 95\% CI. (Old)}
\centering
\begin{tabular}[t]{rrr}
\toprule
$\tilde p$ & Condition (D1)/(D3) & Condition (D2) \\
\midrule
1.0 & 0.873 [0.805, 0.919] & 0.873 [0.805, 0.919]\\
0.9 & 0.853 [0.788, 0.901] & 0.837 [0.773, 0.886]\\
0.8 & 0.836 [0.772, 0.885] & 0.816 [0.752, 0.865]\\
0.7 & 0.818 [0.755, 0.868] & 0.797 [0.734, 0.847]\\
0.6 & 0.799 [0.736, 0.849] & 0.783 [0.722, 0.834]\\
0.5 & 0.777 [0.716, 0.828] & 0.767 [0.706, 0.818]\\
0.4 & 0.752 [0.691, 0.804] & 0.745 [0.684, 0.797]\\
0.3 & 0.721 [0.661, 0.774] & 0.686 [0.627, 0.739]\\
0.2 & 0.679 [0.620, 0.732] & 0.611 [0.554, 0.666]\\
0.1 & 0.609 [0.552, 0.664] & 0.387 [0.333, 0.445]\\
\bottomrule
\end{tabular}
\end{table}

Condition (D4.1) and (D4.3) gave almost the same estimates of the lower worst-case bounds with minimal variances while the estimates in Condition (D4.2) were not satisfied.
We could still evaluate the robustness of meta-analytical results according to Conditions (D4.1) and (D4.3).
When we assumed that no more than 50\% studies were unpublished (i.e., $\p\ge 0.5$), the worst-case bounds of the SAUC generally fell in the interval [0.7, 0.9], indicating good diagnostic test accuracy. 
By the sensitivity analysis of the worst-case bounds, one could know all the possible variations in the estimated SAUC caused by the selective publication of studies.

\end{document}